\documentclass[11pt]{article}
\usepackage{graphicx}
\usepackage{subcaption}
\usepackage{amssymb,amsmath}
\usepackage{float}
\usepackage{tcolorbox}
\usepackage{amsthm}
\usepackage{hyperref}
\hypersetup{
	colorlinks=true,
	linkcolor=blue,
	filecolor=magenta,      
	urlcolor=cyan}

\begin{document}  
\title{Circle packing in arbitrary domains}
\author{
  Paolo Amore \\
  \small Facultad de Ciencias, CUICBAS, Universidad de Colima,\\
  \small Bernal D\'{i}az del Castillo 340, Colima, Colima, Mexico \\
  \texttt{paolo@ucol.mx}
  \and
  Damian de la Cruz \\
  \small Facultad de Ciencias, Universidad de Colima,\\
  \small Bernal D\'{i}az del Castillo 340, Colima, Colima, Mexico \\
  \texttt{ddelacruz1@ucol.mx}
  \and
  Valeria Hernandez \\
  \small Facultad de Ciencias, Universidad de Colima,\\
  \small Bernal D\'{i}az del Castillo 340, Colima, Colima, Mexico \\
  \texttt{vhernandez20@ucol.mx}
  \and
  Ian Rincon \\
  \small Facultad de Ciencias, Universidad de Colima,\\
  \small Bernal D\'{i}az del Castillo 340, Colima, Colima, Mexico \\
  \texttt{irincon@ucol.mx}
  \and
  Ulises Zarate \\
  \small Facultad de Ciencias, Universidad de Colima,\\
  \small Bernal D\'{i}az del Castillo 340, Colima, Colima, Mexico \\
  \texttt{uzarate@ucol.mx}
}

\maketitle

\begin{abstract}
We describe an algorithm that allows one to find dense packing configurations of a number of congruent disks in arbitrary domains in two 
or more dimensions. We have applied it to a large class of two dimensional domains such as rectangles, ellipses, crosses, multiply 
connected domains and even to the cardioid. For many of the cases that we have studied no previous result was available. 
The fundamental idea in our approach is the introduction of "image" disks, which allows one to work with a fixed container, thus lifting the 
limitations of the packing algorithms of \cite{Nurmela97,Amore21,Amore23}.
We believe that the extension of our algorithm to three (or higher) dimensional containers (not considered here) can be done straightforwardly.
\end{abstract}

\maketitle

\section{Introduction}
\label{sec:intro}

We study the packing of a number of congruent disks inside a finite two dimensional region: we aim to find configurations of $N$ disks
inside the domain that maximally cover the internal region without ever overlapping (contacts between the disks themselves and with the external border are allowed). The relevant quantity to maximize is the packing fraction, i.e. the ratio between the area covered by the disk and the area of the container.  Solving this class of problem is known to be challenging despite the apparent simplicity and it usually involves a computational approach. 

The archetype of packing problem is the Kepler's conjecture,  regarding the optimal packing of identical spheres in three (unbounded) dimensions, which was only proved by Thomas Hales in recent times~\cite{Hales05}. The optimal packing in two dimensions, on the other hand, has been proved earlier by Thue~\cite{Thue} and Fejes-Toth~\cite{Fejes42}.

For finite regions in two and three dimensions the border of the container introduces a geometrical frustration which has the effect of lowering sensibly the packing fraction with respect to the value predicted in the absence of a border. Only for $N \rightarrow \infty$ ($N$ is the number of disks) the effect of the border becomes negligible and one expects to tend to the value of packing fraction of the infinite space.
 
Here we are mainly interested in the two dimensional version of the problem, although the ideas at the core of our approach also hold in higher dimensions. In two dimensions circle packing has been studied for a limited number of domains which include the square~\cite{Schaer65, Schaer65b,Goldberg70,Goldberg71, Nurmela97, Nurmela99, Nurmela99b, Szabo07, Szabo07b, Markot21, Amore21}, the circle~\cite{Kravitz67, Reis75, Melissen94b, Graham97a, Graham97b, Graham98, Fodor99, Fodor00, Fodor03, Nemeth98, Nemeth99}, the equilateral triangle~\cite{Oler61, Melissen93, Melissen94, Melissen95, Payan97, Graham04, Joos21}, rectangles of different proportions~\cite{Melissen97, Graham03, Graham09, Specht10, Birgin10}, ellipses~\cite{Birgin13} and more recently regular polygons with an arbitrary number of sides~\cite{Amore23, Amore23b}. A variety of shapes has recently also been studied in \cite{Beasley11, Stoyan12, Beasley19, Machchhar17}.

Additionaly the site "packomania"~\cite{SpechtRepo}, curated by E.Specht, provides an extensive collection of dense packing configurations in two and three dimensions (besides to the domains previously specified, one can find there also the case of isosceles right triangle, the semicircle and a circular quadrant, in two dimensions, and the sphere and the cube, in three dimensions).
Another source of interesting material on packing is the site of E. Friedman~\cite{FriedmanRepo}, which also includes packing of objects other than congruent disks.

The paper is organized as follows: in section \ref{sec:algo} we describe our algorithm, in section \ref{sec:appl} we present a number of applications of the algorithm both to previously studied domains, e.g. rectangles of different proportions or ellipses of different eccentricity, and to domains that, to the best of our knowledge, had not been studied before, such as different deformations of the circle, up to the cardioid and crosses with arm of different length; finally in section \ref{sec:concl} we present our conclusions and elaborate on possible directions for future work.

\section{The algorithm}
\label{sec:algo}

Our approach is based on an algorithm originally developed by Nurmela and Ostergard (N\"O) for the square~\cite{Nurmela97}. 
The fundamental idea at the basis of the latter is to describe the congruent disks as point charges, which interact with a repulsive force and arrange themselves in the domain reaching an equilibrium configuration. One considers the energy functional
\begin{equation}
E = \sum_{i=2}^N \sum_{j=1}^{i-1} \left( \frac{\lambda}{r_{ij}^2} \right)^s \ ,
\label{eq:potential}
\end{equation}
where $r_{ij}$ is the euclidean distance between the $i^{th}$ and $j^{th}$ particles, $N$ is the total number of particles and $\lambda$ is a parameter
that should be set to the minimal squared distance between any two particles. Finally, $s > 0$ is a parameter that controls the nature of the potential: for small (large) values of $s$ the potential is long (short) range.

The points are initially randomly scattered inside a square of side $L$ and they are allowed to interact with a long range force ($s$ small), reaching an equilibrium configuration. The minimal distance between any two points is then interpreted as the diameter of a solid disk, $d=2r$, and thus the configuration can be interpreted as a packing  of $N$ disks inside a square of side $L+d$. The relevant quantity to 
study is the {\sl packing fraction}, sometimes also referred to as {\sl packing density}, which is defined as the ratio between the total area covered
by the disks and the area of the container:
\begin{equation}
\rho = \frac{N \pi r^2}{(L+d)^2}
\end{equation}

Thue~\cite{Thue} and Fejes-Toth~\cite{Fejes42} have proved that the maximum value of the packing fraction that can be obtained in the infinite plane is $\rho_{max} = \pi/\sqrt{12}$, which works as an upper bound when one considers packing in a finite region (the border introduces a geometrical frustration in the problem which makes it harder to reach higher densities).

Once the system has reached equilibrium the value of $s$ is increased, so that the system evolves to a new equilibrium 
and a new $\lambda$ is obtained. However the new configuration reached by the algorithm is accepted only if it improves 
the packing fraction. This process is repeated over and over, until $s$ has reached very large values ($s > 10^6$) and the interaction has effectively become a contact interaction between rigid disks.

In ref.~\cite{Amore21} a modification of this algorithm, which adds border repulsion in the first stages of the algorithm, has been proposed to help avoiding overpopulating the border when the potential is long range and an excessive number of charges could be pushed on the border (once a charge ends on the border it will only be able to move on the border, but never to re--enter the domain, given the fact that the resultant of the forces of all other charges points outward). In this way the modified algorithm increases the probability of generating denser configurations.
In ref.~\cite{Amore23} the algorithm of \cite{Amore21} has been adapted to work for arbitrary regular polygons, which include the square and the circle as special cases, obtaining a large number of dense configurations for a variety of regular polygons.  The numerical findings obtained with the algorithms of ref.~\cite{Amore23} have lead in ref.~\cite{Amore23b} to the observation that for regular polygons with a number of sides multiple of $6$ the best (densest) configurations are curved hexagonal (CHP) for $N$ (number of disks) corresponding  to hexagonal numbers ($N= 3 k (k+1)+1$ and $k=1,2,\dots$)~\footnote{CHP had been previously discovered for the circle by Graham and Lubachevsky~\cite{Graham97a} long time ago.}. At sufficiently large $N$  CHP is lost and the configurations typically do not have any symmetry~\footnote{While the initial discovery of {\rm CHP} in regular polygons of ref.~\cite{Amore23b} originated from numerical observations, that paper provides a {\sl deterministic algorithm} that allows to construct {\sl all} the CHP configurations for a given $N$. }.

However the original N\"O algorithm~\cite{Nurmela97} and the later algorithms of \cite{Amore21,Amore23} suffer of an important limitation, that we now explain. In these algorithms the dimension of the disks is not fixed: as a matter of fact one works with point charges which are confined into a domain; these points represent the centers of the disks and the diameter of the disks corresponds to the minimal distance between any two charges. For this reason, the domain in which the disks are contained is obtained by "expanding" the original domain as shown for example in fig.~1 of \cite{Amore21} for the square or in fig.~1 of \cite{Amore23} for the pentagon. In order for the algorithm to work, the internal and external domains have to be related by a scale transformation, which indeed is the case for regular polygons.

\begin{figure}
\begin{center}
\bigskip\bigskip\bigskip
\includegraphics[width=6cm]{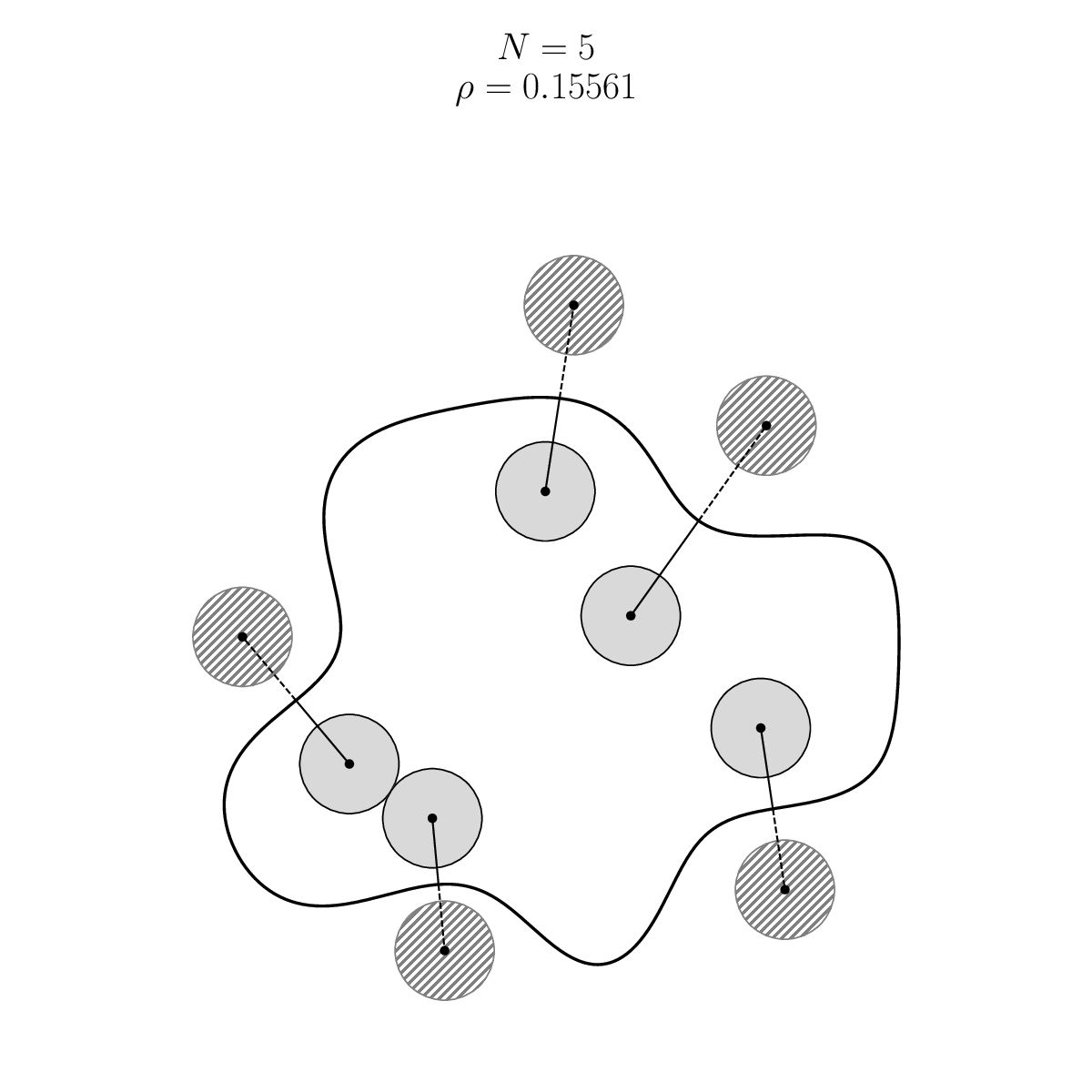}\hspace{0.cm}
\includegraphics[width=6cm]{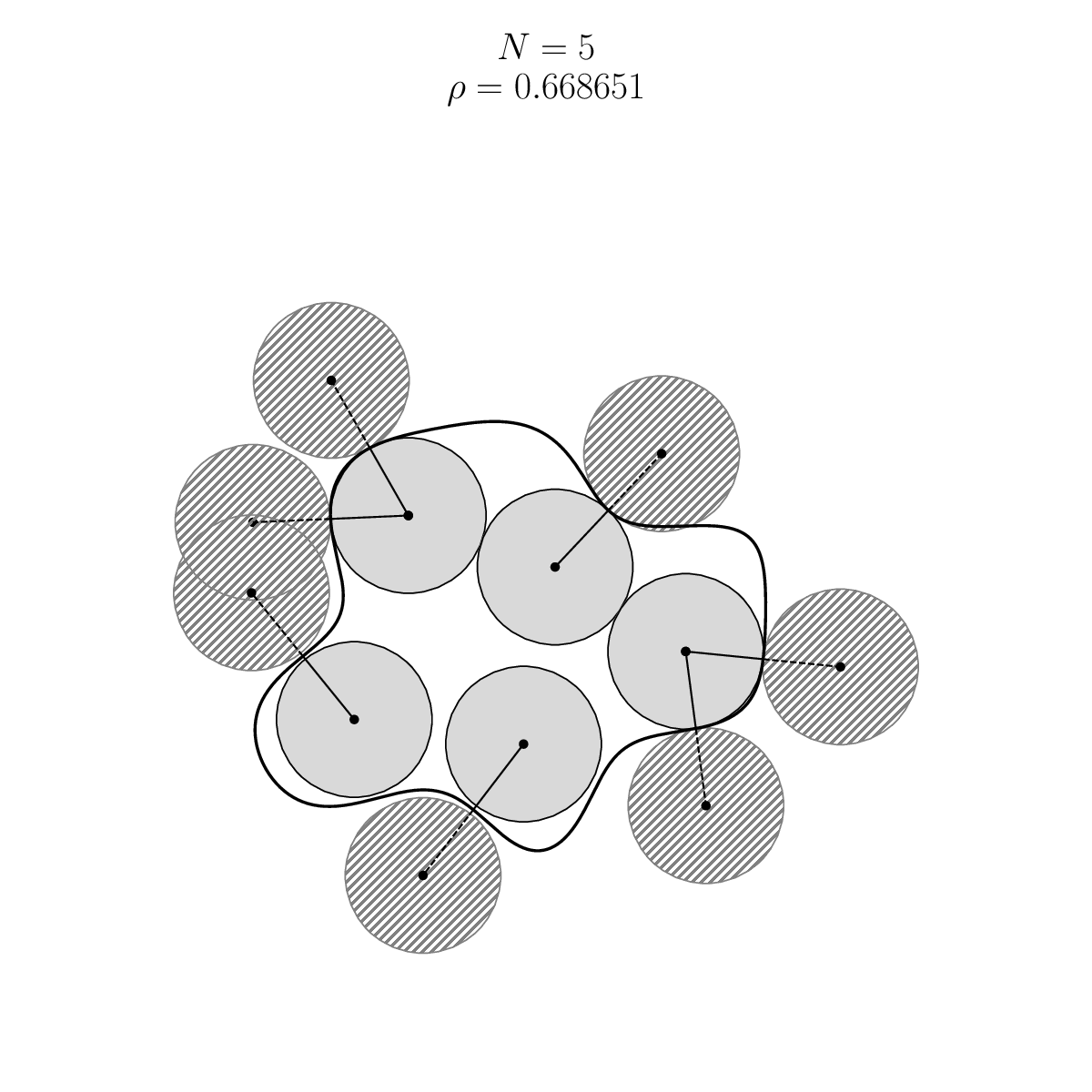}\hspace{1cm}
\caption{Example of disks images in a packing configurations in a container of arbitrary shape.}
\label{Fig_images}
\end{center}
\end{figure}

It is straightforward to see that this property is not fulfilled for more general cases, such as for example rectangles and ellipses (with the exception of the limit cases of square and circle). To understand why this happens, consider a rectangle of sides $L_x$ and $L_y$ and assume that in it one has disks of radius $r$. The centers of the disk are contained inside this rectangle, while the disks themselves are contained into a rectangle of sides $L_x+2r$ and $L_y+2r$. As the algorithm proceeds the value of the radius increases and, as a result, the proportions of the outer rectangle are changing, which completely invalidates the algorithm (if the shape of the domain is changing one cannot make a meaningful comparison between two configurations with the same number of points but two different $r$). 

At first sight it may appear that everything is lost but we have managed to find a clever modification that allows one to immediately adapt the algorithm to work for an arbitrary domain. The fundamental idea at the basis of our method is inspired by the well--known method of images in electrodynamics and it is illustrated in Fig.~\ref{Fig_images}, that we use here to explain the basic idea.

In the left plot of the figure, $5$ points are initially randomly scattered inside a container of arbitrary shape: one can associate to each of the points  at least one {\sl image} point outside the domain, along a direction which is perpendicular to the border and at minimal distance (if there are more points at a minimal distance the number of images increases);  the size of the disks is not determined, as before, by the shortest distance between any two internal points, but by the shortest distance overall, including the distance between a charge and its image. The algorithm is otherwise unmodified and works as we have explained above: the identification of the image charges and the subsequent calculation of the minimal distance are the only new ingredients in the algorithm and bring some additional computational complexity into the problem but they allow at the same time to deal with very different regions with the same approach. In this way, one is working with a fixed container and, as the algorithm advances, the size of the disks is increasing but the container is kept the same. The right plot shows the final configuration at the end of the algorithm (notice that  in this case two of the internal charges have a double image each).

The modification that we have introduced allows one to deal with domains of completely arbitrary nature (observe for instance that the region in Fig.~\ref{Fig_images} is not convex). In fact, as we will discuss in the next section, our algorithm can be also applied to multiply connected domains, to concave domains and even to domains with a border singularity, such as the cardioid.  

It is important to remark that our algorithm requires to solve a constrained optimization problem at each stage, i.e. for each value of $s$, as the disks need to be fully contained in the domain. However it is possible to convert the problem to an {\sl unconstrained} one by introducing an appropriate parametrization of the points, which enforces the constrain automatically and reduces the computational complexity of the problem by a large amount.

 We parametrize a point inside the domain as
\begin{equation}
{\bf P} = (x,y) = \mathcal{R}(t) \cdot C(u)  ,
\label{eq_para} 
\end{equation}
where $C(u)$ is periodic function of $u$ with period $2\pi$ which parametrizes the border of the domain and $\mathcal{R}(t)$ is a function typically bounded between 
$0$ and $1$  (the form of this function is not unique, but we have usually chosen $\mathcal{R}(t) = \sin^2(t)$). Notice that by choosing $\mathcal{R}(t)$ appropriately one can also describe multiply connected regions.

Remarkably, the use of image charges in our algorithm allows one to deal also with {\sl concave} domains on the same footing as with convex domains by simply specifying the appropriate $C(u)$ inside eq.~(\ref{eq_para})  (it must be said that the location of the image charges for a concave region may be computationally heavier, reflecting in a slower process).

\begin{figure}
\begin{center}
\bigskip
\includegraphics[width=3cm]{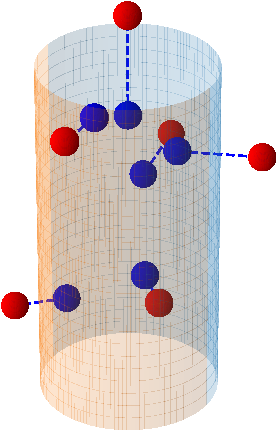}
\caption{Example of application of the algorithm to a finite cylinder in three dimensions (the image spheres are in red).}
\label{Fig_3D}
\end{center}
\end{figure}

Our algorithm can be applied also to higher dimensional problems with minor modifications (not considered in this paper): for the case of a container in three dimensions, for example, one should find the image charges by tracing the lines passing through a given charge normal to the border surface (see Fig.~\ref{Fig_3D}). 
The only expected difference when going from two to higher dimension is an increase in the computational difficulty due to the task of finding the images. A similar argument would apply in dimensions $D > 3$, where however it would be more difficult to use one's intuition as a guide.

Before we turn our attention to the applications it is worth discussing an additional step that can be added to our algorithm: the original algorithm of Nurmela and Ostergard was modified in refs.~\cite{Amore21, Amore23} to allow for border repulsion in the early stages of the minimization, when the interaction is long range. As a matter of fact the configuration will not evolve to a densely packed one, if an excessive number of charges is deposited on the border at this time. A simple way of incorporating this feature in the present approach is to attribute a larger charge to the image points in the first steps of the algorithm and gradually letting it evolve to the custom value when the interaction becomes short range.

\section{Applications}
\label{sec:appl}

In this section we want to discuss some applications of our algorithm to domains of different shape, particularly rectangles, crosses and ellipses of variable proportions, but also the cardioid and  multiply connected domains. Our explorations here are mainly focused on illustrating how our algorithm works, rather than on performing an in depth search of global minima. In this respect there could be cases where the configuration reached is not yet optimal.

\begin{figure}
\centering
\includegraphics[width=0.8\textwidth]{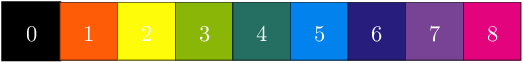}
\caption{Color scheme used for diagrams: the numbers within each square represents the number of contacts of disks or number of sides of Voronoi cells.}
\label{fig_color_scheme}
\end{figure}

In plotting the configurations obtained or the Voronoi diagrams associated to those we will adopt the color scheme described in Fig.~\ref{fig_color_scheme}.

For reasons of convenience we will show in this paper only a limited number of figures representing the configurations that we have calculated: a large collection of figures will however be found in the supplemental material.

\subsection{Rectangles of different proportions}
\label{sub:rect}

As we have mentioned in the introduction, the packing inside rectangles (including the square) has been studied in depth before (see the references previously cited).
We don't pretend here to perform an exhaustive exploration, but rather to concentrate on special values of $N$ which correspond either to perfect squares ($N=25,36,49,64$)~\footnote{In the case of perfect squares it was found in  \cite{Nurmela97} that $N=49$ is the first value where square packing ceases to be optimal and an irregular, denser configuration emerges.} or to special cases where the packing fraction is known to be particularly large ($N=120$)~\cite{Nurmela99b}.

Our goal is to observe what happens  in each of the cases above when one of the sides of the square is increased.  Contrary to eq.~(\ref{eq_para}), here we parametrize a point inside the rectangle of sides $a$ and $1$ as
\begin{equation}
\begin{cases}
x=a\sin{u_1}\\
y=\sin{u_2}
\end{cases}
\end{equation}
Using the algorithm described in the previous section, we have obtained numerical results for many rectangles gradually increasing $a$ with a small stepsize of $0.01$.

\begin{figure}
\centering
\includegraphics[width=5.5cm]{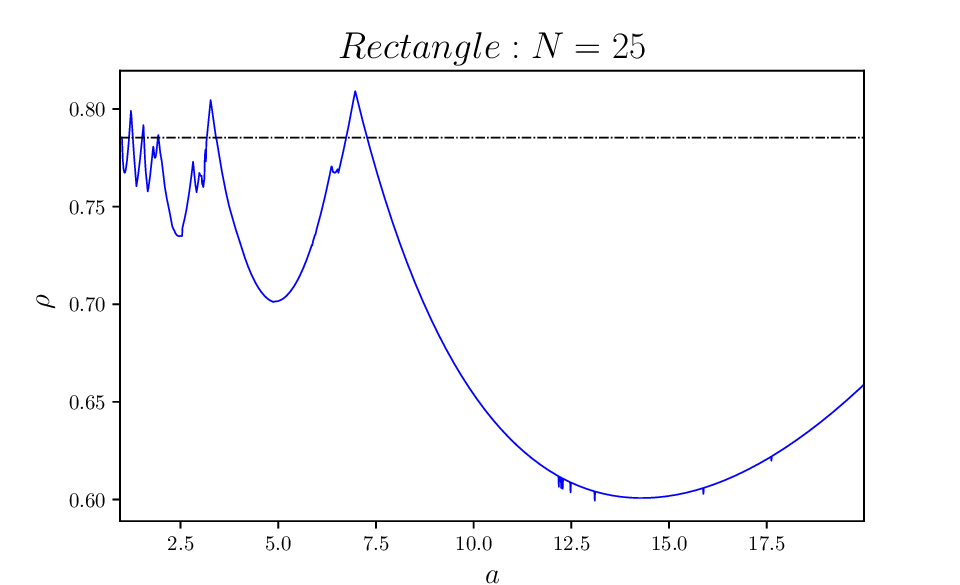} \hspace{1cm}
\includegraphics[width=5.5cm]{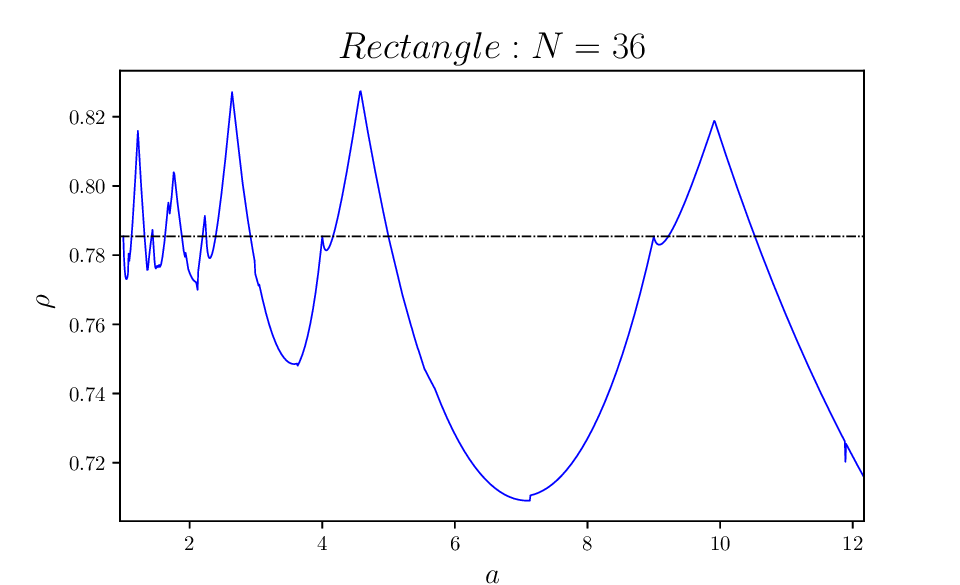} \\
\includegraphics[width=5.5cm]{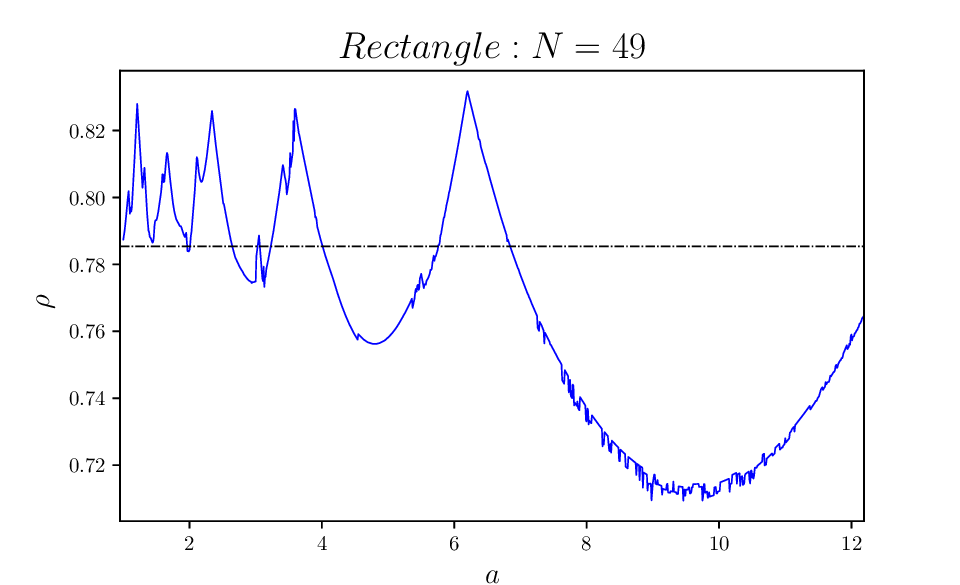}\hspace{1cm}
\includegraphics[width=5.5cm]{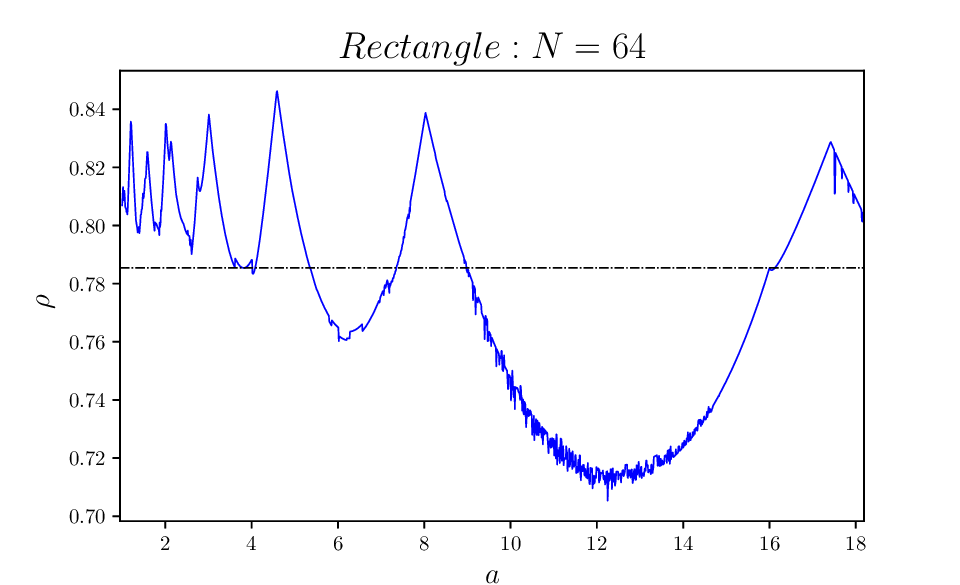} \\
\includegraphics[width=7cm]{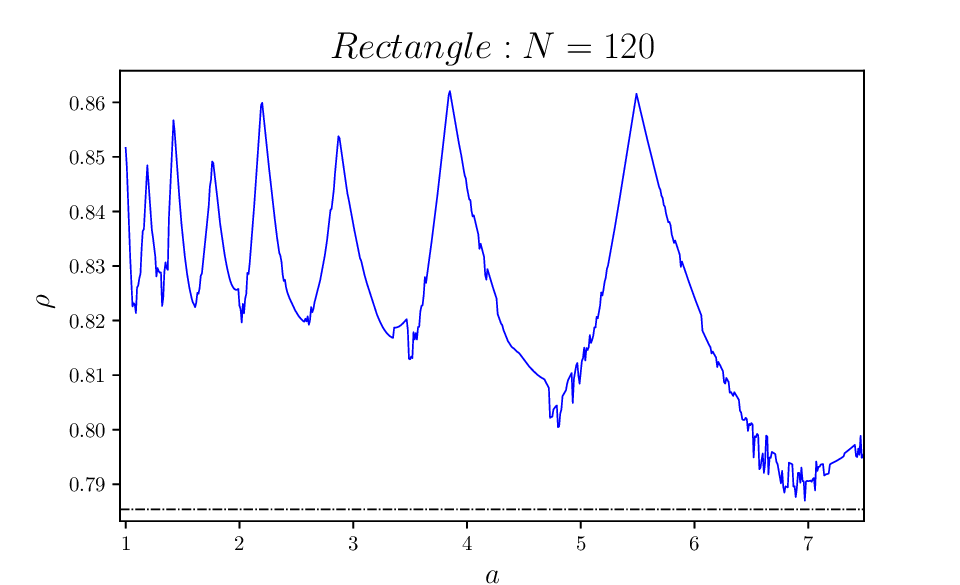}
\caption{Plots of densities for different proportions and number of disks in the rectangle. The black dashed line corresponds to the density of square packing.}
	\label{fig:dens_rect}
\end{figure}

In Fig.~\ref{fig:dens_rect} we plot the packing fraction obtained for these cases as a function of $a$. A common feature in all the plots is the presence of 
multiple peaks and  valleys. The peaks correspond to particularly advantageous configurations where a larger number of disks can be accommodated into a hexagonal packing. The thin dashed horizontal line corresponds to the value $\pi/4$, i.e. the packing fraction for square packing.

The structure of the configurations corresponding to the maxima of the curves in fig.~\ref{fig:dens_rect} will in general differ sensibly from case to case although 
configurations with large packing fraction will contain larger regions of circles packed in a hexagonal arrangement. This is the case, for example, of the plot for $N=36$ in fig.~\ref{fig:dens_rect}:  as a matter of fact the disks are arranged in a perfect square packing  for the square  ($\rho = \pi/4$), but at larger ratio new larger maxima are found ($a=4$ is a neat exception since in this case the disks arrange again in a perfect square packing).

\begin{figure}
\centering
\includegraphics[width=7cm]{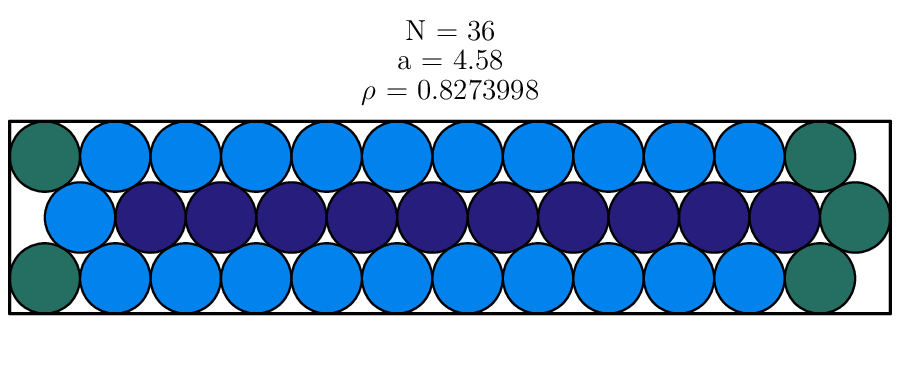}
\caption{Configuration corresponding to the largest maximum of Fig.~\ref{fig:dens_rect} for $N=36$.}
\label{fig:rectangle_36}
\end{figure}

In Figure \ref{fig:rectangle_36} we show the configuration  corresponding to the largest maximum of Fig.~\ref{fig:dens_rect} for $N=36$: in this case a large density is achieved by arranging the disks into a perfect hexagonal packing.

For the case of a rectangle it is not difficult to guess the irrational value of the ratio at which certain dense configurations occur: 
\begin{align}
a=
\begin{cases}
		\frac{2N+l-1}{l(2+(l-1)\sqrt{3})}\quad , \quad l \hspace{5pt} odd \quad\wedge\quad (N+(l-1)/2) \hspace{5pt} mod \hspace{5pt} l = 0 \\
		\frac{2N+l}{l(2+(l-1)\sqrt{3})}\quad , \quad l \hspace{5pt} even \quad\wedge\quad (N+l/2) \hspace{5pt} mod \hspace{5pt} l = 0
	 \\
\frac{2N+l}{l(2+(l-1)\sqrt{3})} \quad, \quad N \hspace{5pt} mod \hspace{5pt} l = 0
\end{cases}
\label{eq:ratios_rect}
\end{align}
corresponding to the density
\begin{align}
\rho=\begin{cases}
		\frac{Nl\pi}{(2N+l-1)(2+(l-1)\sqrt{3})}\quad , \quad l \hspace{5pt} odd \quad\wedge\quad (N+(l-1)/2) \hspace{5pt} mod \hspace{5pt} l = 0 \\
		\frac{Nl\pi}{(2N+l)(2+(l-1)\sqrt{3})}\quad , \quad l \hspace{5pt} even \quad\wedge\quad (N+l/2) \hspace{5pt} mod \hspace{5pt} l = 0 \\
	    \frac{Nl\pi}{(2N+l)(2+(l-1)\sqrt{3})} \quad, \quad N \hspace{5pt} mod \hspace{5pt} l = 0
	\end{cases}
	\label{eq:dens_rect}
\end{align}

\begin{figure}
\centering
\includegraphics[width=7cm]{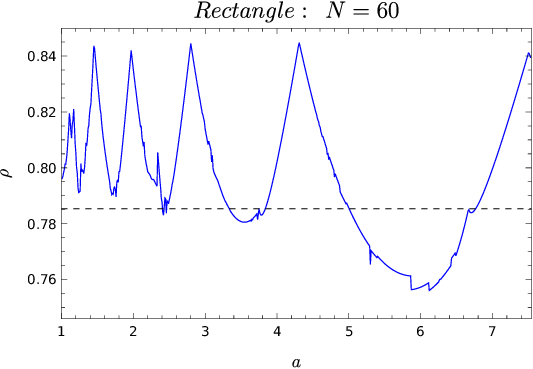}
\caption{Plots of densities for different proportions and number of disks in the rectangle. The black dashed line corresponds to the density of square packing.}
\label{fig:dens_rect2}
\end{figure}

According to the formulas above for $N=60$ one should expect very dense configurations in the rectangle at specific values of the ratio; this is actually seen in 
Fig.~\ref{fig:dens_rect2}, where the peaks correspond to $a$ $=$ $\frac{16}{2+7 \sqrt{3}}$, $\frac{1}{\frac{1}{9}+\frac{1}{\sqrt{3}}}$, $\frac{21}{2+5 \sqrt{3}}$, $\frac{25}{2+4 \sqrt{3}}$, $\frac{31}{2+3 \sqrt{3}}$,  $\frac{41}{4} \left(\sqrt{3}-1\right)$ (the maxima at $a= 2.34$, $3.75$ and $6.67$ do not correspond to this formula). Notice that at these peaks the density is drastically larger than in the square (at special values of the ratio one can also have less dense  maxima, corresponding to square packing).

\subsection{Cross of different proportions}
\label{sub:cross}

The second class of example that we consider is a symmetric cross with arms of variable length; we call $a$ the length of the arms, with $a=0$ corresponding to the unit square. 

Leaving religious considerations apart, the cross is interesting because it is an example of concave domain (which is technically more difficult to study) and to the best of our knowledge circle packing inside a cross is an uncharted territory (for a complete different problem, the spectrum of the laplacian on a domain, 
it is know however that the infinite cross has localized states, which are normally classified as "bound states in the continuum" or BIC). 

\begin{figure}
	\centering
	\begin{subfigure}{0.45\textwidth}
		\centering
		\includegraphics[width=\textwidth]{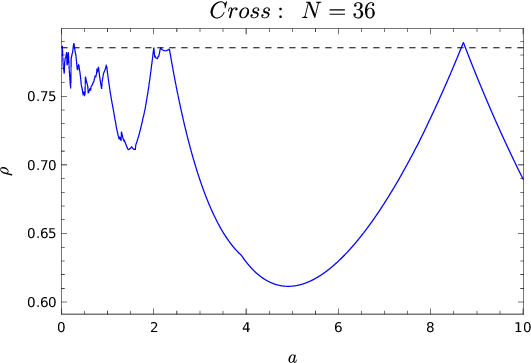}
	\end{subfigure}\hfill
	\begin{subfigure}{0.45\textwidth}
		\centering
		\includegraphics[width=\textwidth]{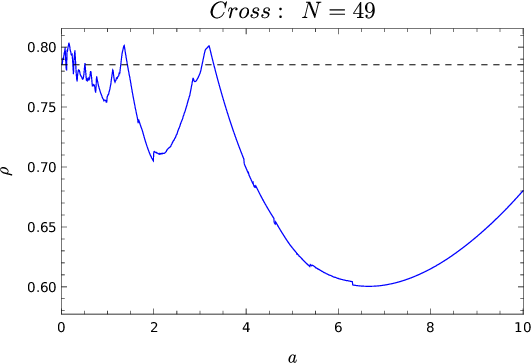}
	\end{subfigure}
	\vspace{0.5cm} 
	\begin{subfigure}{0.45\textwidth}
		\centering
		\includegraphics[width=\textwidth]{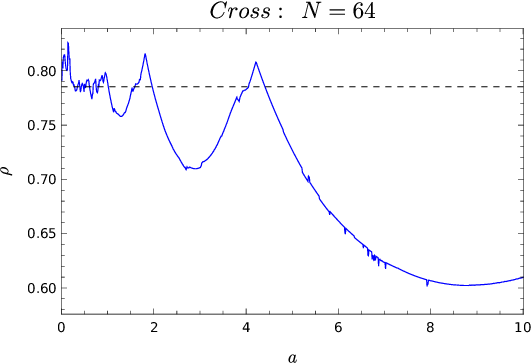}
	\end{subfigure}
	\caption{Plots of densities for different proportions and number of disks in the cross. The black dashed line corresponds to the density of square packing.}
	\label{fig:dens_cross}
\end{figure}

\begin{figure}[h]
\centering
\includegraphics[width=4cm]{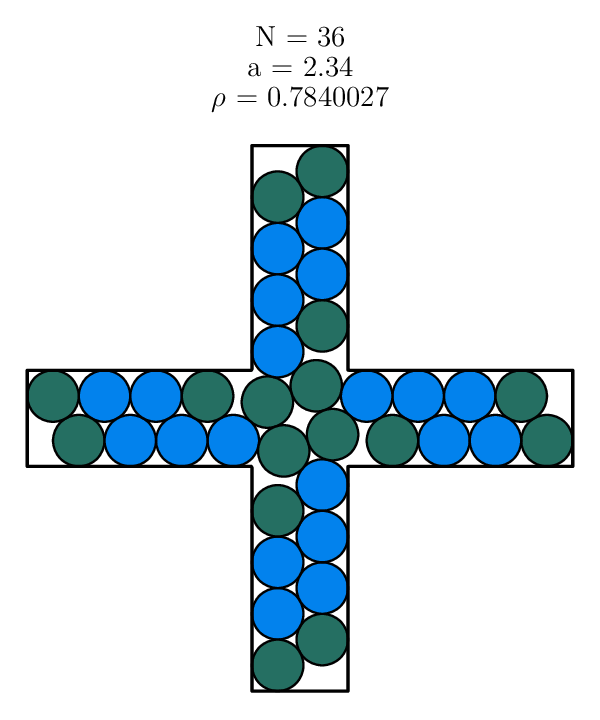} \hspace{1cm}
\includegraphics[width=4cm]{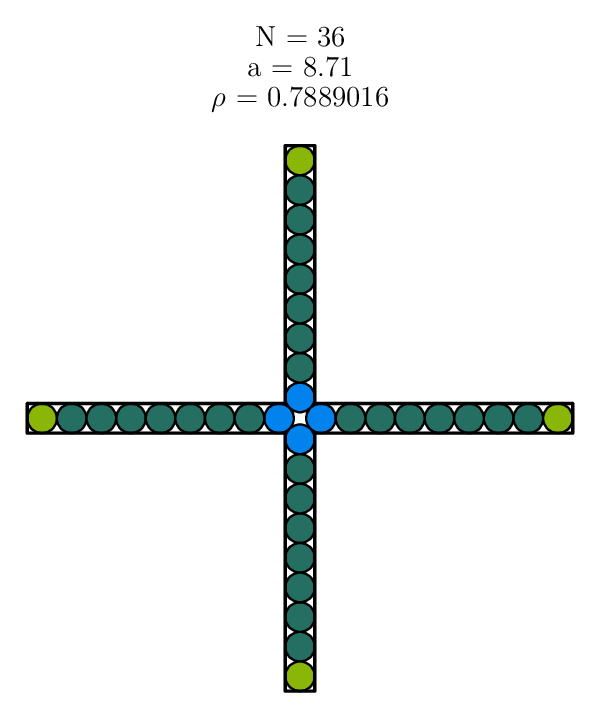}
\caption{Two configurations corresponding to maxima of the density for $N=36$.}
\label{fig:cross_36}
\end{figure}

We have carried out numerical calculations with our algorithm for crosses with arms of varying length and with $N=36$, $49$ and $64$. The main reason for considering these values of $N$ is the fact that a square with $N=36$ disks is the last example in which square packing in the square is optimal, whereas $N=49$ and $64$ are the first two cases where square packing is non optimal. The behavior of the density as function of $a$ is plotted in fig.~\ref{fig:dens_cross}.
Notice that the last maximum in the plot corresponding to $N=36$ has a density slightly larger than $\pi/4$ (the density of square packing): this is entirely due to the efficient packing close to the center of the cross (see the right plot of fig.~\ref{fig:cross_36}). For $a \approx 2.34$ the optimal configuration corresponds to a simmetric packing  containing both square and hexagonal arrangements, with a density slightly below $\pi/4$.

\begin{figure}[h]
\begin{center}
\includegraphics[width=4cm]{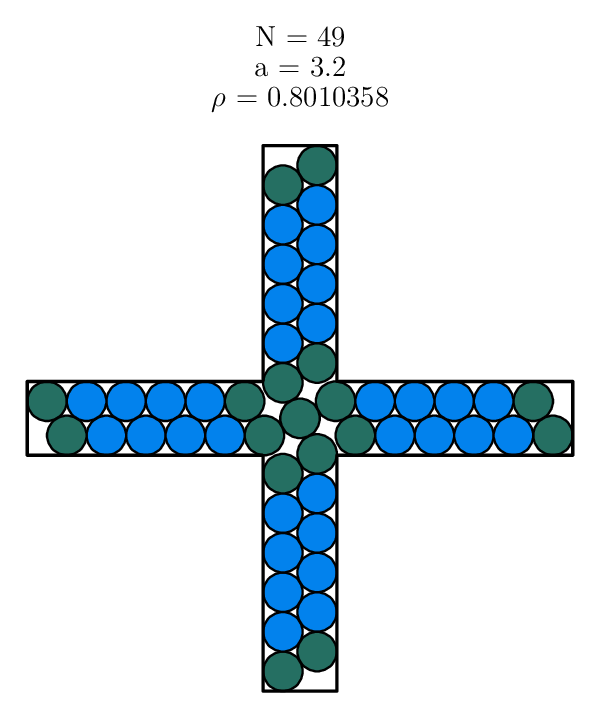} \hspace{0.3cm}
\includegraphics[width=4cm]{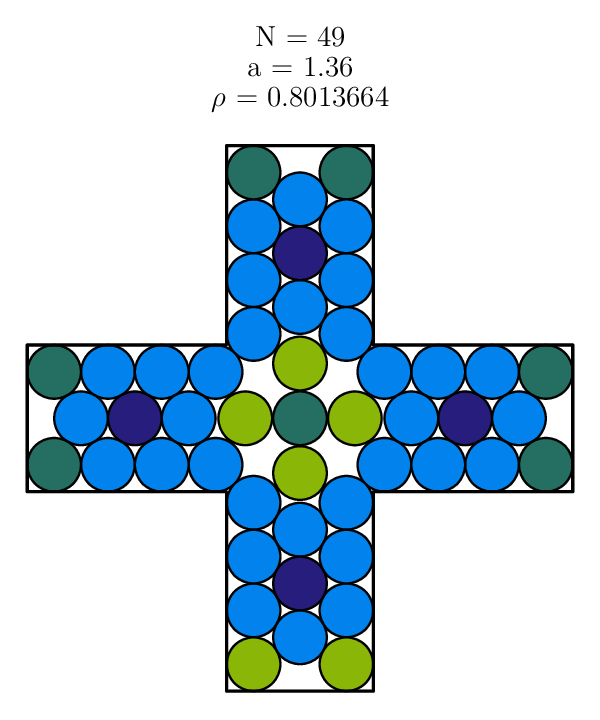}
\caption{Two configurations corresponding to maxima of the density for $N=49$.}
\label{fig:cross_49}
\end{center}
\end{figure}

Similar configurations are also found at different $N$: in fig.~\ref{fig:cross_49}, for example, we show two configurations corresponding to maxima of the packing fraction for $N=49$.

\subsection{Ellipses of different proportions}
\label{sub:ell}
 
The next case that we pretend to study is the ellipse that we parametrize in terms of
\begin{equation}
C(u) = (a\cos{u},\sin{u})
\end{equation}
In this case $a \geq 1$ the length of the major axis, which controls the eccentricity of the ellipse ($a=1$ is the circle).

\begin{figure}
	\centering
	\begin{subfigure}{0.45\textwidth}
		\centering
		\includegraphics[width=\textwidth]{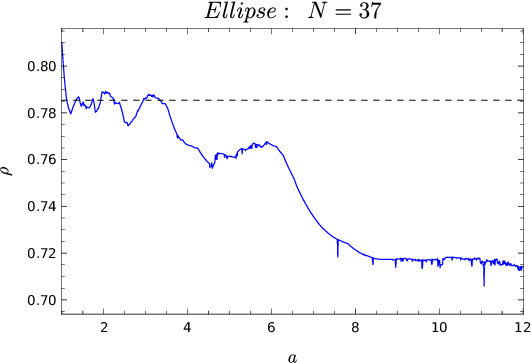}
	\end{subfigure}\hfill
	\begin{subfigure}{0.45\textwidth}
		\centering
		\includegraphics[width=\textwidth]{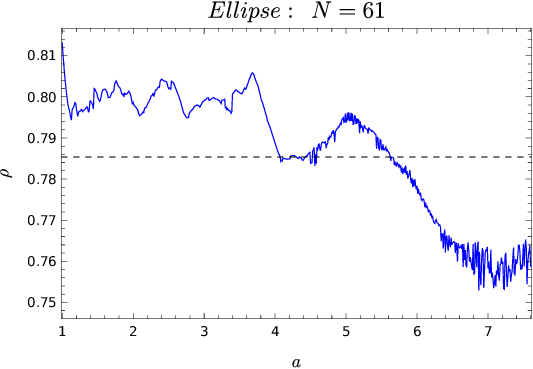}
	\end{subfigure}
	\vspace{0.5cm} 
	\begin{subfigure}{0.45\textwidth}
		\centering
		\includegraphics[width=\textwidth]{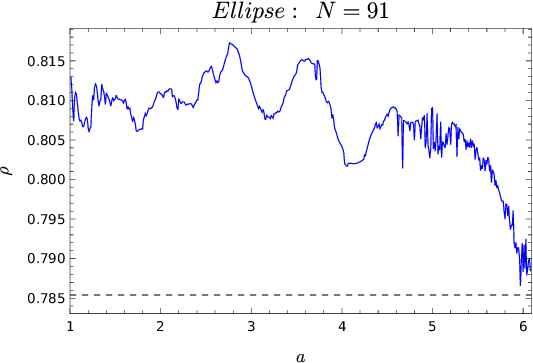}
	\end{subfigure}
	\caption{Plots of densities for different proportions and number of disks in the ellipse. The black dashed line corresponds to the density of square packing.}
	\label{fig:dens_ellipse}
\end{figure}

\begin{figure}
\centering\includegraphics[width=8cm]{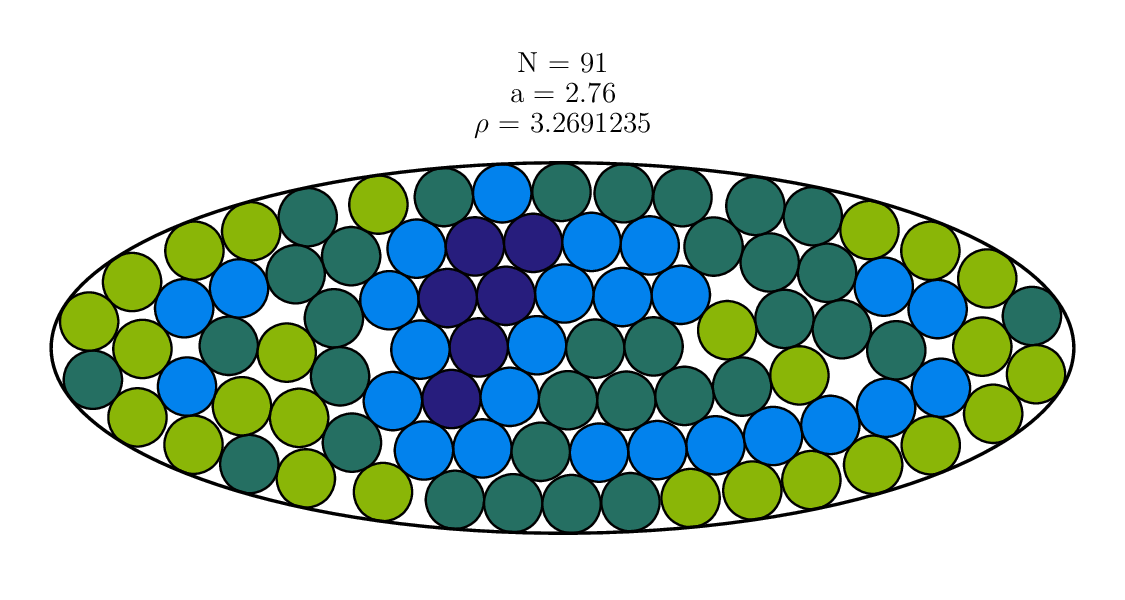}
\caption{Dense packing configuration in the ellipse for $N=91$.}
\label{fig:ellipse_91}
\end{figure}

Packing in the ellipse has been studied previously by Birgin et al. in \cite{Birgin13}; much in the spirit of our previous examples, here we want to see what happens to a particular configuration inside the circle when the domain is deformed to an ellipse of given length $a$. We have chosen to consider the cases $N=37$, $61$ and $91$ which correspond to CHP inside the circle; in Fig.~\ref{fig:dens_ellipse} we plot the packing fraction as a function $a$ for these $N$, by performing numerical simulation starting from $a$ to some larger value with small steps $\delta a =0.01$. Also in this case we observe the presence of multiple maxima of the packing fraction. It is important to notice that only in the case of $N=91$ there are maxima that lie above the CHP value, the largest one corresponding to $a \approx 2.76$ (see fig.~\ref{fig:ellipse_91}).

\subsection{Continuous deformation of a circle into a cardioid}
\label{suc:card}
For this case we use the following parametrization which has constant area of $\pi$:
\begin{equation}
	C(u) = \sqrt{\frac{2}{2+a^2}}\cdot (\cos{u}-a\sin^2u,(1+a\cos{u})\sin{u})
\end{equation}

\begin{figure}
	\centering
	\begin{subfigure}{0.45\textwidth}
		\centering
		\includegraphics[width=\textwidth]{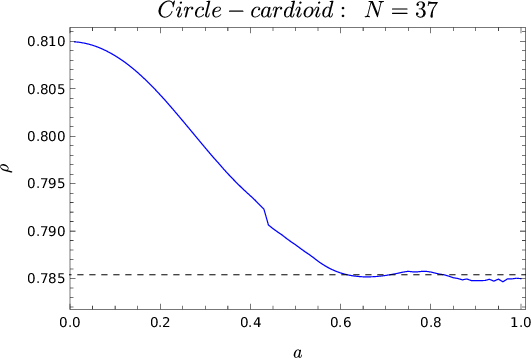}
	\end{subfigure}\hfill
	\begin{subfigure}{0.45\textwidth}
		\centering
		\includegraphics[width=\textwidth]{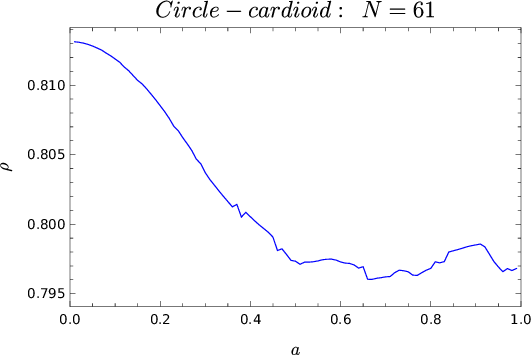}
	\end{subfigure}
	\vspace{0.5cm} 
	\begin{subfigure}{0.45\textwidth}
		\centering
		\includegraphics[width=\textwidth]{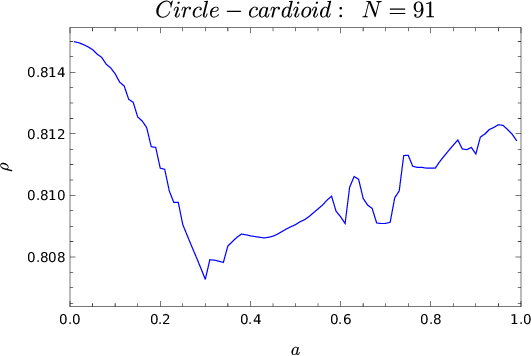}
	\end{subfigure}
	\caption{Plots of densities for different proportions and number of disks in the circle-cardioid. The black dashed line corresponds to the density of square packing.}
	\label{fig:dens_circCard}
\end{figure}

In Fig.~\ref{fig:dens_circCard} we plot the packing fraction as a function of $a$ for $N=37$, $61$ and $91$ (corresponding to CHP configurations in the circle). As before we notice the presence of maxima for $a>1$, although we haven't found any configuration denser than those of the circle.

The case of a cardioid ($a=1$) deserves special attention since it is a concave domain and with a border singularity at the cusp. For the different problem of $N$ equal  charges on a curved line (Thomson problem) studied in ref.~\cite{Amore19}, it was found that the cusp has a dramatic effect on the distribution of charges as $N$ grows. This justifies the curiosity of knowing whether the cusp has a sizable effect also on the packing structure. 

We have conducted a number of numerical experiments with our algorithm trying to find the optimal configurations of congruent disks in a cardioid for several values of $N$ (the largest being $N=2000$, which however is certainly not very accurate since obtained with a single trial). 

\begin{figure}
	\centering
	\includegraphics[width=0.7\textwidth]{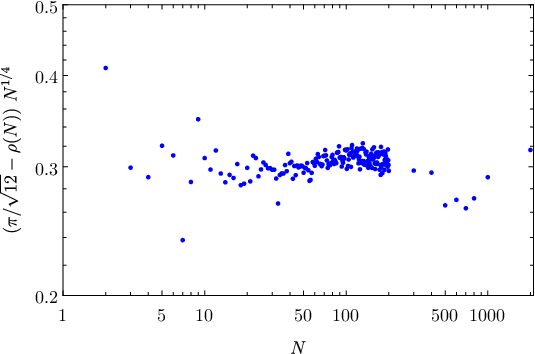}
	\caption{Plot of densities for different number of disks in the cardioid of area $\pi$. The black dashed line corresponds to the density of square packing.}
	\label{fig:dens_card}
\end{figure}

In Fig.~\ref{fig:dens_card} we plot the quantity $(\pi/\sqrt{12} - \rho) N^{1/4}$, which seems to suggest that in this case the packing fraction has a behavior
\begin{equation}
\rho \approx \frac{\pi}{\sqrt{12}}  - \frac{\kappa}{N^{1/4}}  
\end{equation}
with $\kappa \approx 0.3$. Of course, more solid conclusions should be reached after studying configurations with larger $N$ (and improving the present results performing a larger number of trials).


\begin{figure}
	\centering
	\includegraphics[width=4cm]{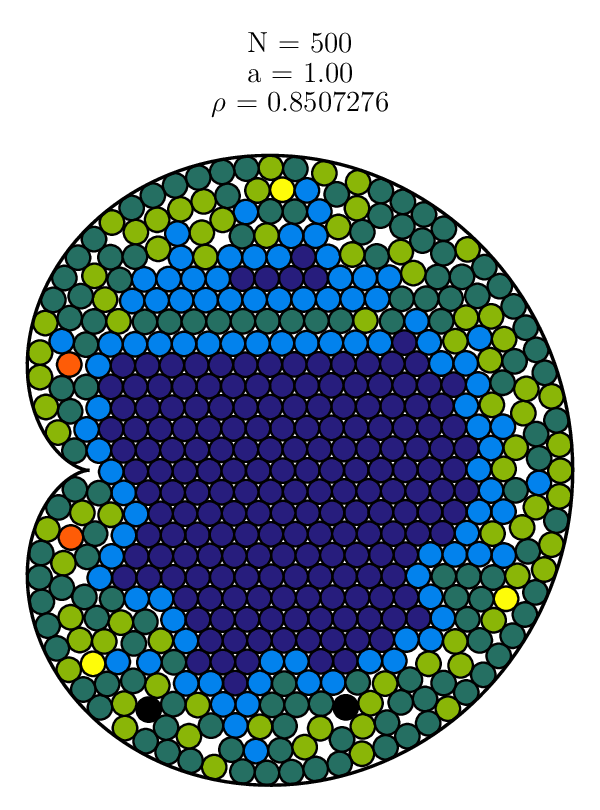} \hspace{1cm}
	\includegraphics[width=4cm]{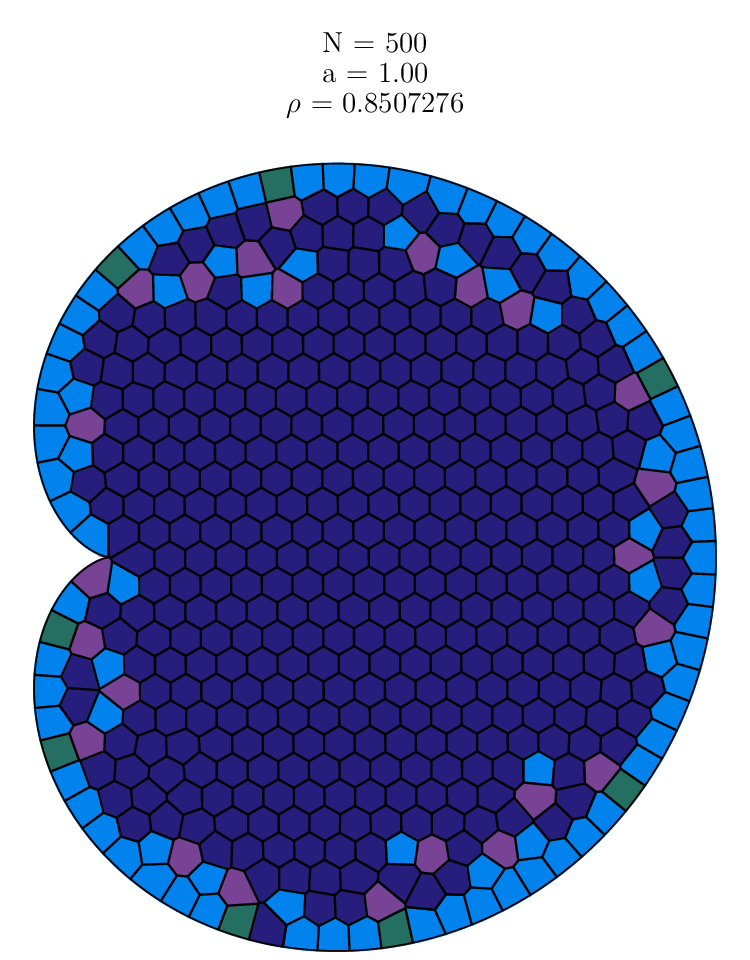}
	\caption{Configuration of $500$ disks inside a cardioid.}
	\label{fig:cardioid_500}
\end{figure}

In Fig.~\ref{fig:cardioid_500} we plot the densest configuration of $500$ disks in the cardioid that we have found with repeated applications of our algorithm.
As expected, the central region of the domain is entirely covered by hexagonal Voronoi cells,  with pentagonal and heptagonal cells appearing close to the border.

\subsection{Multiply connected domains}
\label{sub:multi}

The last example that we want to consider is domain with a hole (multiply connected). Both the domain and the hole may have arbitrary shapes, although we will focus only a circular domain with a circular concentric hole (some other cases can be found in the supplemental material ~\cite{supp1}).

In Fig.~\ref{fig:annulus} we consider the special case of an annulus with inner radius $a$ for $N=90$ disks and plot the packing fraction as function of $a$. This case is particularly interesting because at a specific value of $a$ corresponding to the radius of the disks ($a\approx 0.09$) the resulting configuration correspond to  curved hexagonal packing in a circle, found long time ago by Graham and Lubachevsky~\cite{Graham97a}, after removing the central disk (actually such configuration is also found for $a < 0.09$). This configuration is shown in the upper left plot of Fig.~\ref{fig:annulus_2}; the remaining configurations in the figure correspond symmetric configurations found at selected maxima of the packing fraction (see Fig.~\ref{fig:annulus}). Eventually, for $a$ sufficiently large, all the disks are pushed to the border and progressively shrinked as $a\rightarrow 1^{-}$.

Another aspect to notice is the fact that the total topological charge vanishes in this case, just as it happens in the infinite plane. One way to realize this would be considering Voronoi diagrams with a  sea of hexagonal cells, separated from the border by pentagonal cells. The Voronoi diagrams for the configurations of Fig.~\ref{fig:annulus_2} are displayed in Fig.~\ref{fig:annulus_3}.
Notice that if the border of a Voronoi cell has both a part of outer and inner border, its topological charge is now lowered by two: for example, the configurations for to $3 \leq N\leq 13$ only 
have cells with four sides (see \cite{supp1}), but since these cells both share the inner and outer border their  topological charge is $0$, and the corresponding topological charge of the whole domain is also vanishing, as it should be.

\begin{figure}
\centering
\includegraphics[width=8cm]{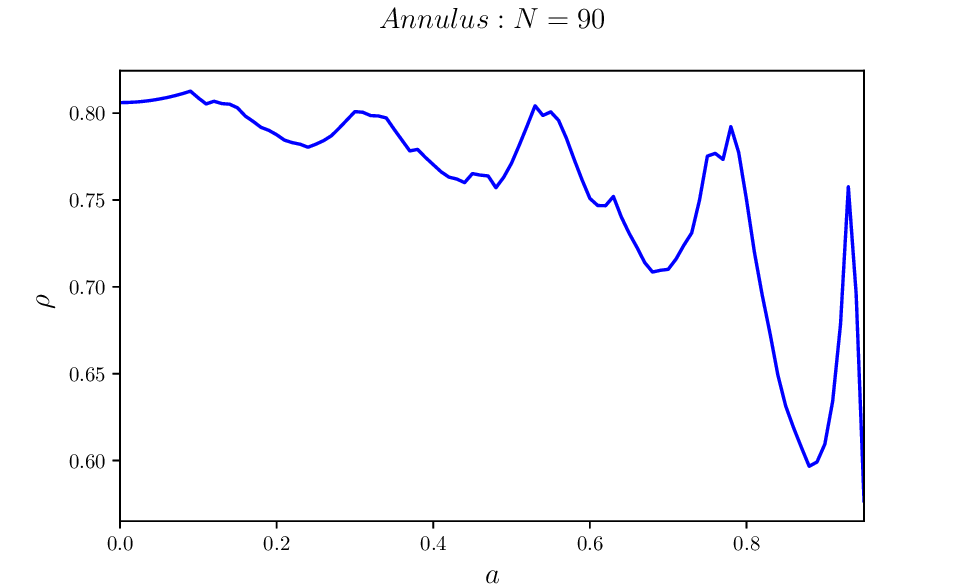} 
\caption{Packing fraction in an annulus as a function of the inner radius for $N=90$.}
\label{fig:annulus}
\end{figure}

\begin{figure}
\centering
\includegraphics[width=4cm]{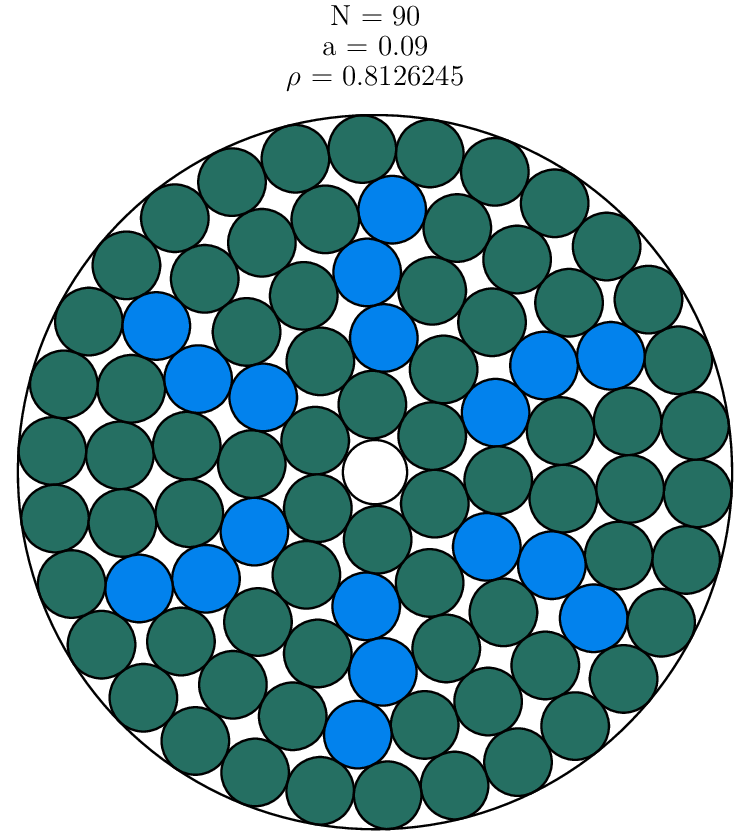}  \hspace{0.5cm}
\includegraphics[width=4cm]{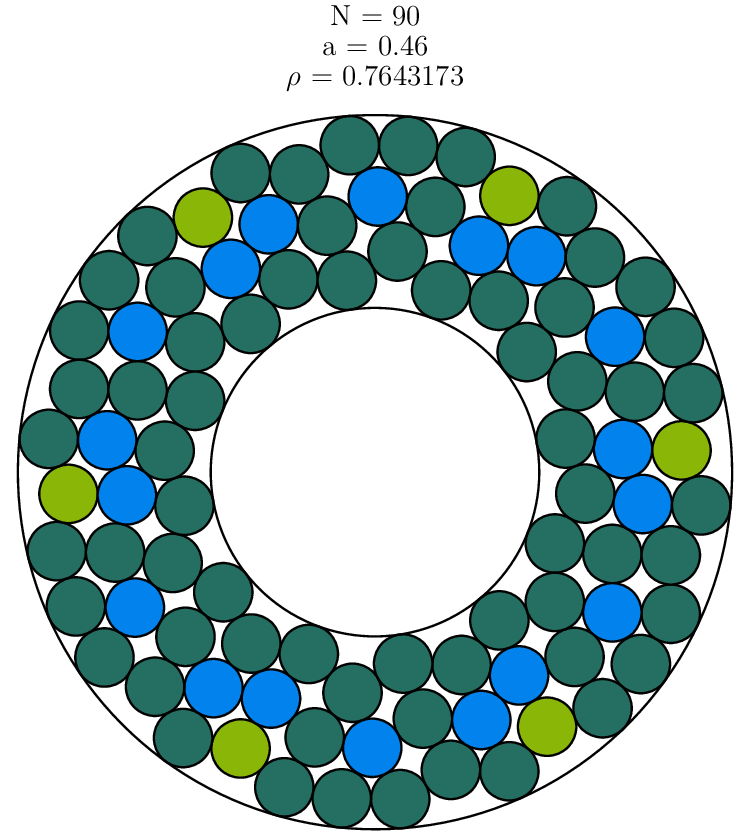} \\
\includegraphics[width=4cm]{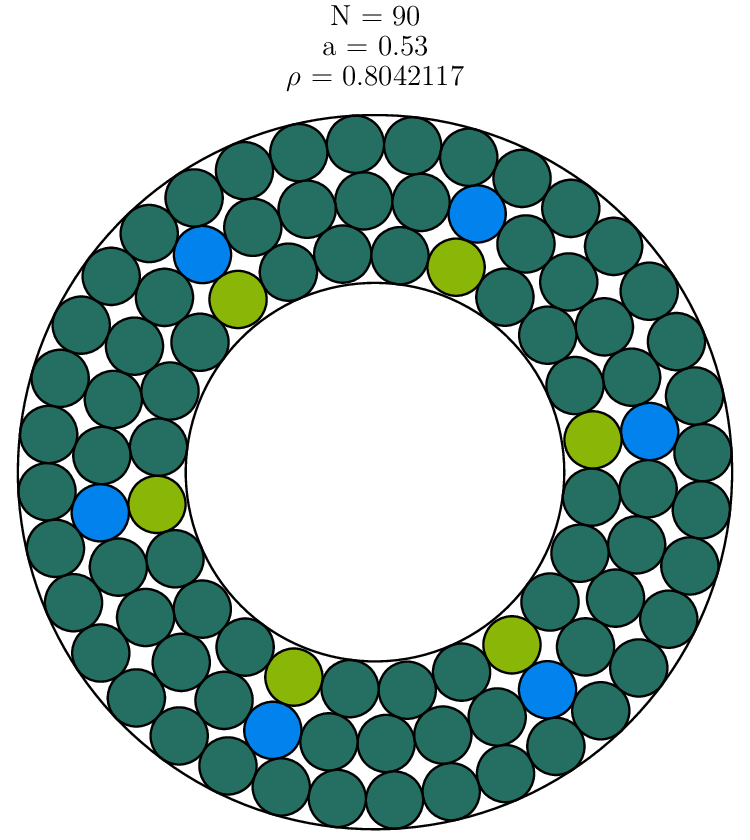}  \hspace{0.5cm}
\includegraphics[width=4cm]{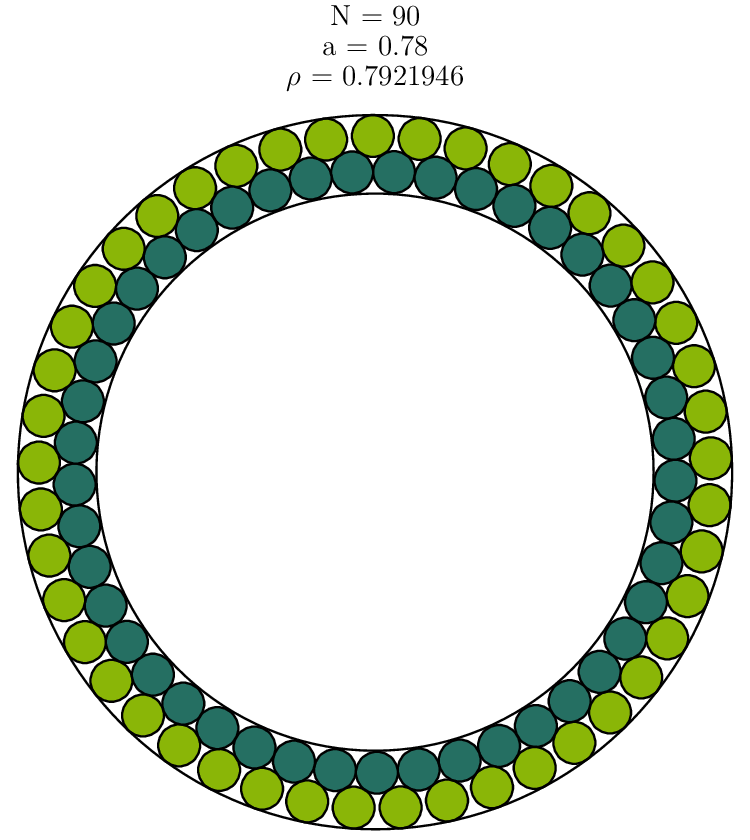} \\
\caption{Symmetric configurations inside an annulus with $N=90$ disks for different values of $a$.}
\label{fig:annulus_2}
\end{figure}

\begin{figure}
\centering
\includegraphics[width=4cm]{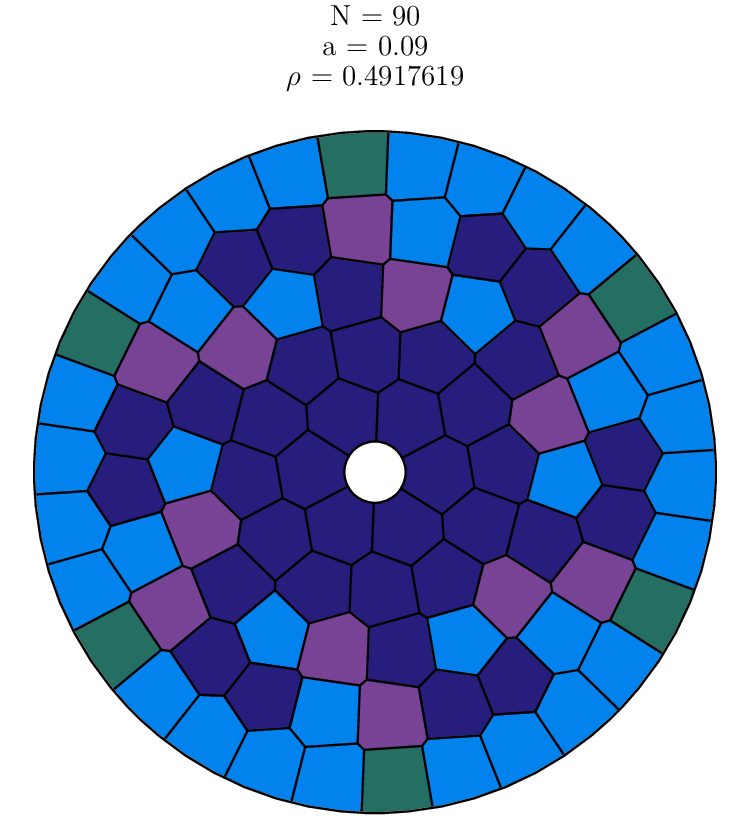}  \hspace{0.5cm}
\includegraphics[width=4cm]{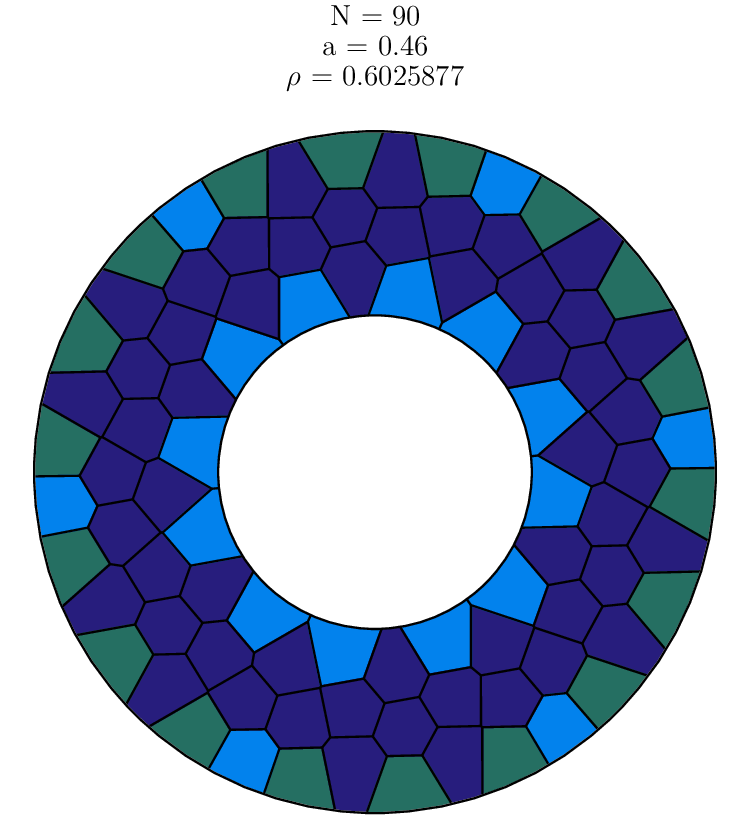} \\
\includegraphics[width=4cm]{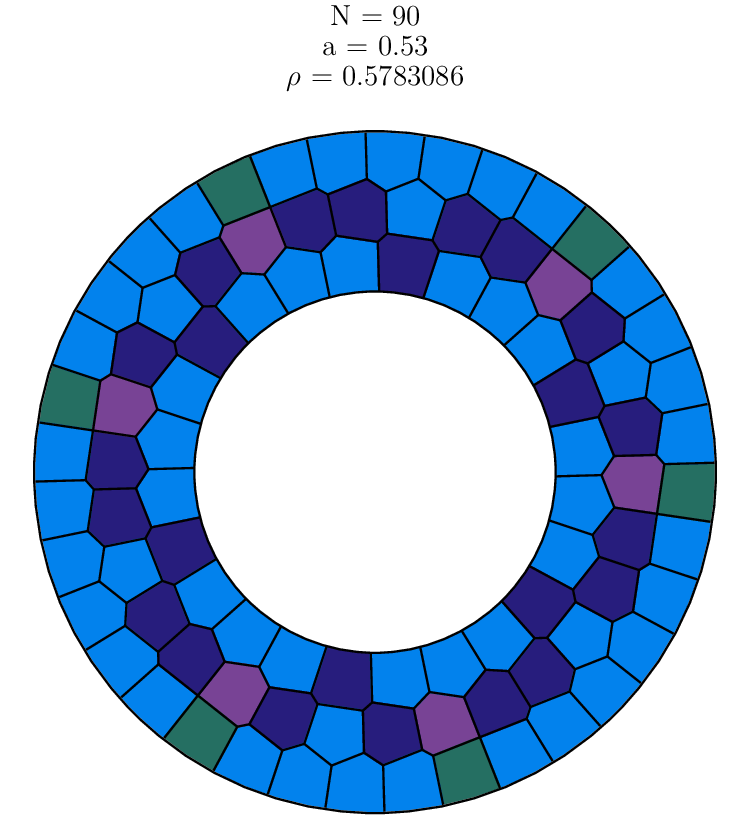}  \hspace{0.5cm}
\includegraphics[width=4cm]{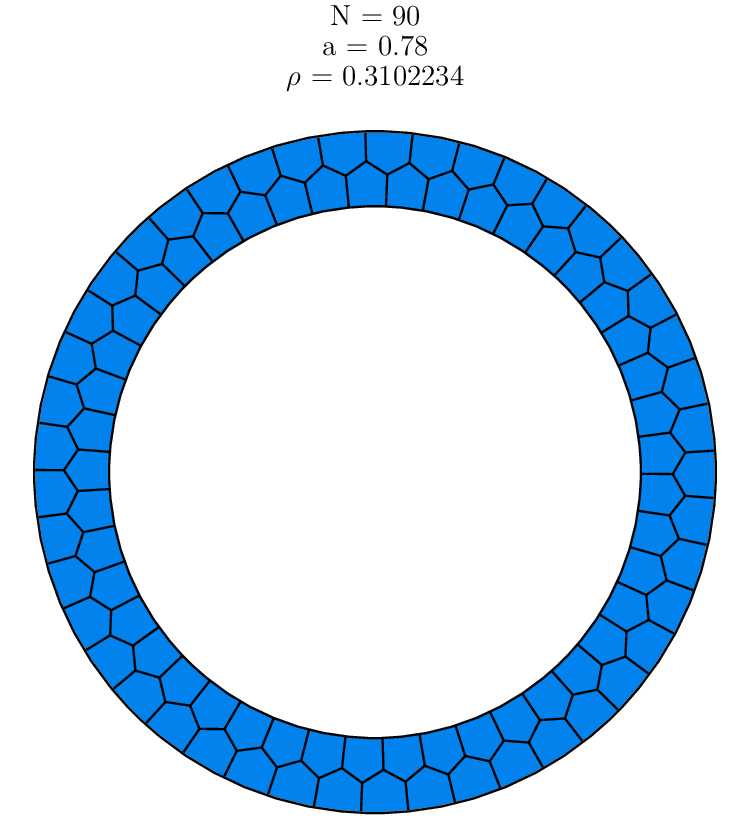} \\
\caption{Voronoi diagrams of symmetric configurations inside an annulus with $N=90$ disks for different values of $a$.}
\label{fig:annulus_3}
\end{figure}

\section{Conclusions}
\label{sec:concl}

In this paper we have devised an algorithm to find dense packing configurations of a number of congruent disks inside an arbitrary domain.
Although we have focused only on two dimensional regions, our algorithm applies to domains in any dimension with minor modifications: in fact the  essential step in its implementation  is identifying the image charges, which can be done for problems of arbitrary dimensions and domains of arbitrary shape.

We have applied our method to study a large variety of domains, many of which had never been considered before. In some cases we have found that possible global maxima of the packing fraction correspond to highly symmetric solutions (particularly for the cross and the annulus). The case of  the cardioid is also interesting because of the presence of a border singularity (the cusp), which seems to affect the large $N$ behavior of the packing fraction. 


Future work with more realistic applications should consider three dimensional regions, particularly packing in tubular containers which has relevant applications in physics (and biophysics).   We believe that it would be possible to study numerically a larger class of containers (particularly cylinders with an arbitrary section) due to the flexibility of our algorithm. Finally, the application to higher dimensional domains ($D \geq 4$), while less relevant for physical applications, is also possible and may constitute an interesting direction of research.

\section*{Acknowledgements}
The plots in this paper have been plotted using Matplotlib for python~\cite{matplot} and Mathematica~\cite{wolfram} with {\rm MaTeX} \cite{szhorvat}. 
Numerical calculations have been carried out using python ~\cite{python} and numba ~\cite{numba}. 
The research of P.A. was supported by Sistema Nacional de Investigadores (M\'exico). 
P. A. would like to acknowledge support from the Isaac Newton Satellite program for attending the Geompack workshop,
where the  work was completed.

\end{document}